Spin-Orbit Torque Efficiency in Compensated Ferrimagnetic Cobalt-Terbium Alloys

Joseph Finley[1] and Luqiao Liu[1]

[1]Department of Electrical Engineering and Computer Science,
Massachusetts Institute of Technology, Cambridge, Massachusetts 02139, USA

Despite the potential advantages of information storage in antiferromagnetically coupled materials, it remains unclear whether one can control the magnetic moment orientation efficiently because of the cancelled magnetic moment. Here, we report spin-orbit torque induced magnetization switching of ferrimagnetic $Co_{1-x}Tb_x$ films with perpendicular magnetic anisotropy. Current induced switching is demonstrated in all of the studied film compositions, including those near the magnetization compensation point. The spin-orbit torque induced effective field is further quantified in the domain wall motion regime. A divergent behavior that scales with the inverse of magnetic moment is confirmed close to the compensation point, which is consistent with angular momentum conservation. Moreover, we also quantify the Dzyaloshinskii-Moriya interaction energy in the $Ta/Co_{1-x}Tb_x$ system and we find that the energy density increases as a function of the Tb concentration. The demonstrated spin-orbit torque switching, in combination with the fast magnetic dynamics and minimal net magnetization of ferrimagnetic alloys, promises spintronic devices that are faster and with higher density than traditional ferromagnetic systems.

I. Introduction

There has been great interest recently in using antiferromagnetically coupled materials as opposed to ferromagnetic materials (FM) to store information. Compared with FM, antiferromagnetically coupled systems exhibit fast dynamics, as well as immunities against perturbations from external magnetic fields, potentially enabling spintronic devices with higher speed and density [1,2]. Rare earth (RE)– transition metal (TM) ferrimagnetic alloys are one potential candidate material for realizing such devices. Inside RE-TM alloys, the moments of TM elements (such as Fe, Co, Ni) and the RE elements (e.g., Gd, Tb, Ho, etc) can be aligned with anti-parallel orientations due to the exchange interaction between the f and d electrons [3]. By varying the relative concentrations of the two species, one can reach compensation points where the net magnetic moment or angular momentum goes to zero [4–7]. Moreover, because of the different origins of magnetism in the two species, transport related properties are dominated by TM in these alloys, providing a way to read out the magnetic state even in a compensated system. In this work, we show that by utilizing the current induced spin-orbit torque, one can switch magnetic moments in Ta/$Co_{1-x}Tb_x$ bilayer films. Particularly, we found that effective fields generated from the spin-orbit torque scaled with the inverse of magnetization and reached maximum when the composition approaches the magnetic compensation point. The large effective spin-orbit torque and the previously demonstrated fast dynamics [8,9] in these ferrimagnetic systems provide a promising platform for high speed spintronic applications.

II. Characterization of Magnetic Properties

Spin-orbit torque (SOT) originates from spin-orbit interaction (SOI) induced spin generation in the bulk (i.e., the spin Hall effect) [10,11] or the surface (i.e., the Rashba-Edelstein effect) [12,13] of solid materials. So far, SOTs have proven to be an efficient method of controlling the ferromagnetic state of nanoscale devices [14–17]. Recently, it was demonstrated that one can utilize SOT to switch TM-dominant CoFeTb ferrimagnets with perpendicular magnetic anisotropy (PMA) [18]. It is therefore interesting to ask what the relationship is between the chemical composition of RE-TM alloys and the SOT efficiency, and whether or not one can switch compensated ferrimagnets using SOT. To answer these questions, we grew a series of Ta(5)/Co$_{1-x}$Tb$_x$(t)/Ru(2) (thickness in nm) films using magnetron sputtering. The Co$_{1-x}$Tb$_x$ alloys were deposited by co-sputtering Co and Tb sources with different sputtering powers. The concentration of Tb $x$ was calculated from the deposition rates and varied between 0.1 and 0.3, while the layer thickness $t$ ranges from 1.7 nm to 2.6 nm. The magnetic properties of the deposited films were examined using vibrating sample magnetometry (VSM), and PMA was observed for all samples [Fig. 1 (a)]. Furthermore, the magnetic moment goes through zero and the coercive fields reach their maximum around $x \approx 0.22$, which is consistent with the room temperature magnetic moment compensation point $x_{cM}$ reported earlier [6,7]. The dependence of the magnetic moment on the Tb concentration is summarized in Fig. 1(c), which agrees well with the trend line calculated by assuming an anti-parallel alignment between the Co and Tb moment (dashed lines). The samples were then patterned into Hall bars with dimensions 4 x 44 μm$^2$ [Fig. 1(d)]. The anomalous Hall resistance ($R_{AH}$) vs magnetic field $H$ curves measured from these samples are plotted in Fig. 1(b) and summarized in Fig. 1(e). The polarity of the hysteresis loops changes sign across $x_{cM}$, consistent with previous studies where $R_{AH}$ is shown to be dominated by the momentum of the Co sublattice [19,20].

III. Current-Induced Switching

Fig. 2(a) illustrates the current-induced magnetic switching for the series of samples. During these measurements, in-plane magnetic fields of ±2000 Oe were applied in the current flowing direction [$y$ axis in Fig. 1 (d)]. Previous studies showed that an in-plane field is necessary to ensure deterministic magnetic switching of PMA films, as it can break the symmetry between two equivalent final states [21–26]. As shown in Fig. 2(a), the current-induced switching shows opposite polarities under the positive and negative applied fields, consistent with the model of SOT induced switching [26]. Furthermore, under the same in-plane field, the switching polarity changes sign as the samples go from being Co-dominant to Tb-dominant. This phenomenon can be explained by considering a macrospin model as shown in Fig. 2(b). In SOT switching, the Slonczewski torque [27] is proportional to $\hat{m} \times (\hat{\sigma} \times \hat{m})$, where $\hat{m}$ is the unit vector along the magnetic moment direction and $\hat{\sigma}$ is the orientation of electron spins generated from SOI [along the $\hat{x}$ direction in Fig. 1(d)]. Because the torque is an even function of the local magnetic moment $\hat{m}$, effects from both sublattices in the RE-TM alloy add constructively [1,28]. At equilibrium positions, $\hat{m}$ and $\hat{\sigma}$ are perpendicular to each other, and it is usually convenient to use an effective field [22] $H_{ST} \propto \hat{\sigma} \times \hat{m}$ to analyze the SOT effect on magnetic switching. The equilibrium position of $\hat{m}$ can then be determined by balancing the anisotropy field $H_{an}$, the applied in-plane field $H_y$, and $H_{ST}$. As shown in Fig. 2(b), when $H_y$ and $\hat{\sigma}$ are given by the illustrated directions, the final orientation of the Co sublattice magnetic moment will be close to the $+\hat{z}$ direction for the Co dominant sample and close to the $-\hat{z}$ direction for the Tb dominant sample, giving rise to opposite Hall voltages. Based upon this analysis, the current induced switching should exhibit the same polarity change across the compensation point as the magnetic field induced switching. We note that all of the samples follow this rule except for the $Co_{0.77}Tb_{0.23}$ sample, where the field switching data had determined it to be Tb-dominant but the current induced switching corresponds to a Co-dominant sample. A careful study on this sample reveals that this change simply arises from Joule heating induced temperature change. In RE-TM alloys, the magnetic moments of the RE atoms have stronger temperature dependence compared with TM atoms. Consequently, when the temperature increases, a higher RE concentration is necessary to achieve the same magnetic moment compensation point [3–9,29]. By measuring the $R_{AH}$ vs H curves of the $Co_{0.77}Tb_{0.23}$ sample under different applied currents, we found that the polarity of the field induced switching did change sign when the current density is higher than $2 \times 10^7$ A·cm$^{-2}$, suggesting that the sample underwent a (reversible) transition from Tb-dominant to Co-dominant.

IV. Quantitative Determination of Spin-Orbit-Torque Efficiency

The critical current of SOT induced switching in a multi-domain sample is influenced by defect-related factors such as domain nucleation and domain wall (DW) pinning [24]. Therefore, the SOT efficiency *cannot* be simply extracted using the switching current values determined in Fig. 2 (a). To quantify the SOT in our samples, we measured the SOT induced effective field in the DW motion regime by comparing it with the applied perpendicular field, using the approach developed by C. F. Pai et al. [25]. It has been shown that in PMA films with Néel DWs, the Slonczewski term acts on the DW as an effective perpendicular magnetic field and induces DW motion [Fig. 3(g)] [21–25].Therefore, by measuring the current induced shift in the $R_{AH}$ vs $H_z$ curves, one can determine the magnitude of the SOT. Fig. 3 (a) and (b) show typical field induced switching curves for a Co-dominant sample $Co_{0.82}Tb_{0.18}$ and a Tb-dominant sample $Co_{0.75}Tb_{0.25}$. Under the applied current of ±3 mA and in-plane field of 2000 Oe, the centers of hysteresis loops are offset from zero, with opposite values for opposite current directions. The current dependence of the offset fields are summarized in Fig. 3 (c) and (d) for $H_y = 0$ and ±2000 Oe, where a linear relationship between the offset field and the applied current is obtained. In these plots, the ratio between the offset field $H_z^{eff}$ and current density $J_e$ curve represents the efficiency of the SOT at the Ta/RE-TM interface, defined as $\chi \equiv \frac{H_z^{eff}}{J_e}$. $\chi$ as a function of applied $H_y$ for $Co_{0.82}Tb_{0.18}$ and $Co_{0.75}Tb_{0.25}$ samples are plotted in Fig. 3 (e) and (f), respectively. $\chi$ grows linearly in magnitude for small values of $H_y$, until reaching the saturation efficiency $\chi_{sat}$ at a large in-plane field $H_y^{sat}$. The evolution of $\chi$ as a function of $H_y$ comes from the chirality change of the DWs in the sample. It is known that because of the Dzyaloshinskii-Moriya interaction (DMI) mechanism at the heavy metal/magnetic metal interface [30] or inside the bulk of RE-TM alloy [31], stable Néel DW with spontaneous chiralities are formed. Under zero $H_y$, the DWs do not favor either switching polarities, leading to a zero offset field. As $H_y$ increases, the DMI induced effective field $H_{DMI}$ is partially canceled, and DWs start to move in directions that facilitate magnetic switching. $H_y^{sat}$ therefore represents the minimum field that is required to completely overcome $H_{DMI}$ and $\chi_{sat}$ represents the maximum efficiency of the SOT.

Fig. 4 (a) illustrates the dependence of $\chi_{sat}$ on $x$ in $Co_{1-x}Tb_x$ samples. It can be seen that $\chi_{sat}$ diverges near $x_{CM}$, with the largest value occurring for the sample with smallest magnetization. This result is consistent with the spin torque theory, where the ratio between the SOT effective field and applied charge current is $\chi_{sat} = (\pi/2)(\xi\hbar/2e\mu_0 M_s t)$ [25,32]. Here $\xi = J_s/J_e$ represents the effective spin Hall angle, $\frac{\hbar}{2e}J_s$ is the spin current density, $\hbar$ is Planck's constant, $\mu_0$ is the vacuum permeability, and $M_s$ is the saturation magnetization. Note that this model of spin torque is based upon the conservation of total angular momentum. Previously it has been suggested to utilize ferrimagnetic materials with minimized $M_s$ to increase the efficiency of spin torque induced switching [33]. However, it was not verified if an efficient spin absorption could be achieved at the surface of a ferrimagnet material with antiparallel aligned sublattices. Moreover, because of the mixture between the spin angular momentum and orbital angular momentum in RE-TM alloys, there have been debates over the conservation of total angular momentum in these systems [34]. Our experimental results provide clear evidence on the strong efficiency of spin orbit torques in antiferromagnetically coupled materials. Within the experimental accuracy, we found that the effective field from the SOT does follow the simple trend given by $1/M_s t$ [dashed lines in Fig. 4(a)], reflecting total angular momentum conservation. $\xi$ in our samples is determined to be ~0.03, smaller than previously reported values from Ta/magnetic layer devices, possibly due to the relatively smaller spin-mixing conductance at the Ta/CoTb interface [35].

In addition to the magnetic moment compensation point $x_{cM}$, inside RE-TM systems there also exists an angular momentum compensation point $x_{cJ}$ due to the different g factors associated with spin and orbital angular momentum. For our $Co_{1-x}Tb_x$ system, using the g factors of Co (~2.2) and Tb (~1.5) atoms [36,37], along with the relation $J_{Co(Tb)} = M_{Co(Tb)}/\gamma_{Co(Tb)}$, where $\gamma_{Co(Tb)} = -g_{Co(Tb)}\mu_B/\hbar$ ($\mu_B$ being the Bohr magneton, $\gamma_{Co(Tb)}$ the gyromagnetic ratio, and $J_{Co(Tb)}$ the total angular momentum per unit volume), we determine $x_{cJ}$ to be ~17%, which is within the range of the studied samples and lower than $x_{cM}$. Previously it was demonstrated that ultrafast field-driven magnetic dynamics could be excited around $x_{cJ}$ [8,9]. According to Landau-Lifshiz-Gilbert equation of a ferrimagnetic system [27,29,38], the spin torque term leads to $\frac{d\hat{m}}{dt} \sim -\gamma_{eff}\frac{\hbar J_s}{2e\mu_0 M_s t}(\hat{m}\times\hat{\sigma}\times\hat{m})$, where $\gamma_{eff} = (M_{Co}-M_{Tb})/(J_{Co}-J_{Tb})$ is the effective gyromagnetic ratio and $M_S = M_{Co} - M_{Tb}$ [8,9]. When $J_{Co} - J_{Tb}$ approaches zero, the time evolution of $\hat{m}$ diverges if $J_s$ remains finite at $x_{cJ}$. As observed in Fig. 4(a), $\chi_{sat}$ remains roughly unchanged across $x_{cJ}$, suggesting that similar to the field-driven experiment, SOT could also be used as an efficient drive force for achieving fast dynamics at this concentration. Finally, we find the switching polarity keeps the same sign across $x_{cJ}$, differing from the current induced switching of CoGd spin valves studied in Ref. [29], where a switching polarity reversal was observed between $x_{cJ}$ and $x_{cM}$. This difference is due to the presence of different switching mechanisms: in SOT induced switching of PMA films, the two competing torques are the field torque $\gamma_{eff}\mathbf{M} \times \mathbf{H}_{eff}$ and the spin torque. Because the two terms have the same pre-factor $\gamma_{eff}$ and are only functions of $\mathbf{M}$, $\mathbf{H}_{eff}$, and $\hat{\sigma}$, under the same applied $\hat{\sigma}$ and $\mathbf{H}_y$, the orientation of $\hat{m}$ will remain the same (Fig. 2b), regardless of the sign of $\gamma_{eff}$. In contrast, the anti-damping switching of spin valves changes polarity for regions with $\gamma_{eff} < 0$, as explained in Ref. [29].

## V. Measurement of the Dzyaloshinskii-Moriya Interaction Energy

The in-plane field needed for saturating the SOT, $H_y^{sat}$, is plotted against the Tb concentration in Fig. 4(b). First of all, we notice that $H_y^{sat}$ is largest near $x_{cM}$. This result is consistent with fact that the effective DMI field [32] $H_{DMI} = D/M_s t \mu_0 \Delta$, where $D$ is the DMI energy density and $\Delta$ is the DW width, would become divergent when $M_s$ approaches zero. Secondly, $H_y^{sat}$ is generally larger for the Tb-dominant samples than the Co-dominant ones. For example, the sample with the highest Co concentration, $Co_{0.87}Tb_{0.13}$, shows $H_y^{sat}$~100 Oe, which is close to the reported saturation field of Ta/FM stacks [25]. However, in the Tb dominant sample $Co_{0.71}Tb_{0.29}$, which has similar magnetic moment, $H_y^{sat}$ is found to be ~1500 Oe. In the inset of Fig. 4(b) we plot $H_y^{sat} M_s t$, which increases roughly linearly as a function of $x$. By calculating the DW width $\Delta = \sqrt{A/K_u}$ using the determined anisotropy energy $K_u = 6.4 \times 10^4$ J·m$^{-2}$ from Tb- and Co-dominant samples and the reported exchange stiffness [39] A ~ $1.4 \times 10^{-11}$

J/m, we get *D* in the range of 0.05~0.66 pJ/m. The increasing DMI energy with increasing Tb concentration can be explained by the strong spin-orbit coupling and large deviation from the free electron g factor [30] in the Tb atoms. The generation of magnetic textures such as chiral DWs and magnetic skyrmions [40] relies on the competition between the DMI energy and other magnetostatic energies. Therefore, the tunable DMI through chemical composition provides a useful a knob for controlling magnetic phases.

## VI. Conclusion

To summarize, we demonstrated SOT induced switching in $Co_{1-x}Tb_x$ thin films with a wide range of chemical compositions. The effective field from the SOT was found to scale with the inverse of magnetic moment, consistent with the conservation of angular momentum. The high efficiency of SOT at the compensation points as well as the previously demonstrated fast dynamics in these systems makes them highly attractive for high speed spintronic applications. Moreover, we found that the DMI energy density is much larger in samples with high rare earth concentrations, which could provide useful applications in spintronic devices that employ stable magnetic textures.

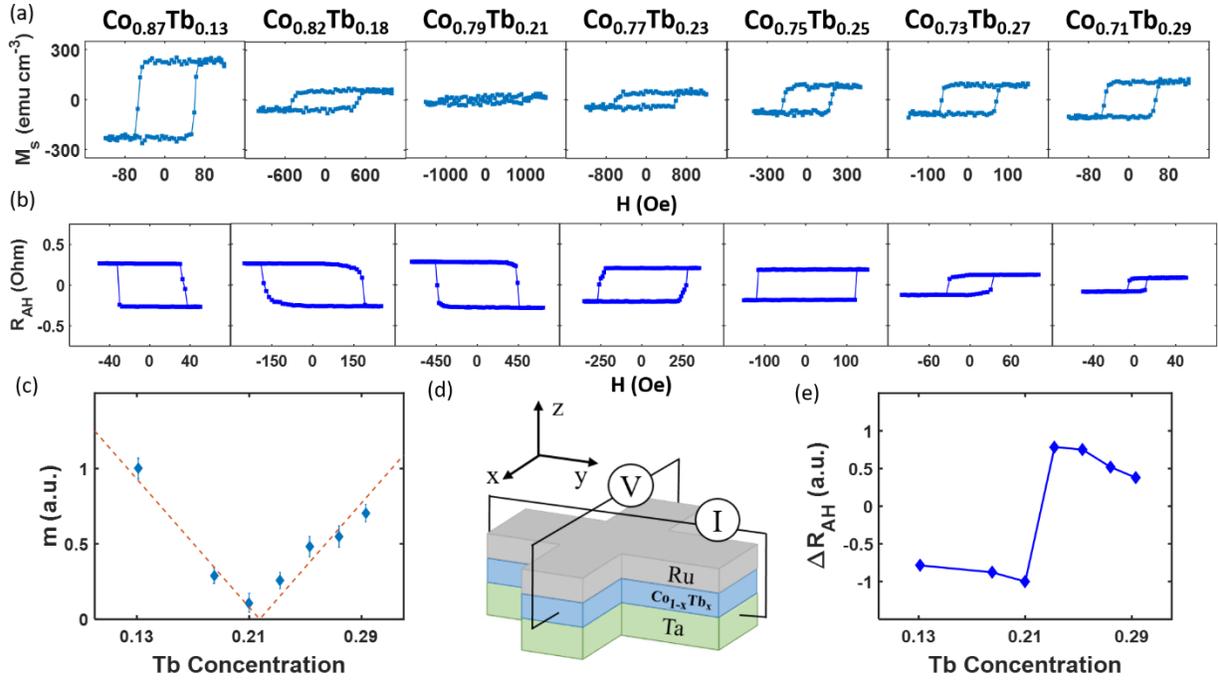

Fig. 1 (a) Out of plane magnetization curves of $Co_{1-x}Tb_x$ films. (b) $R_{AH}$ as a function of perpendicular magnetic field. (c) Magnetic moments of $Co_{1-x}Tb_x$ alloys as a function of Tb concentration. (d) Schematic of the device geometry for $R_{AH}$ measurement. (e) $R_{AH}$ as a function of Tb concentration.

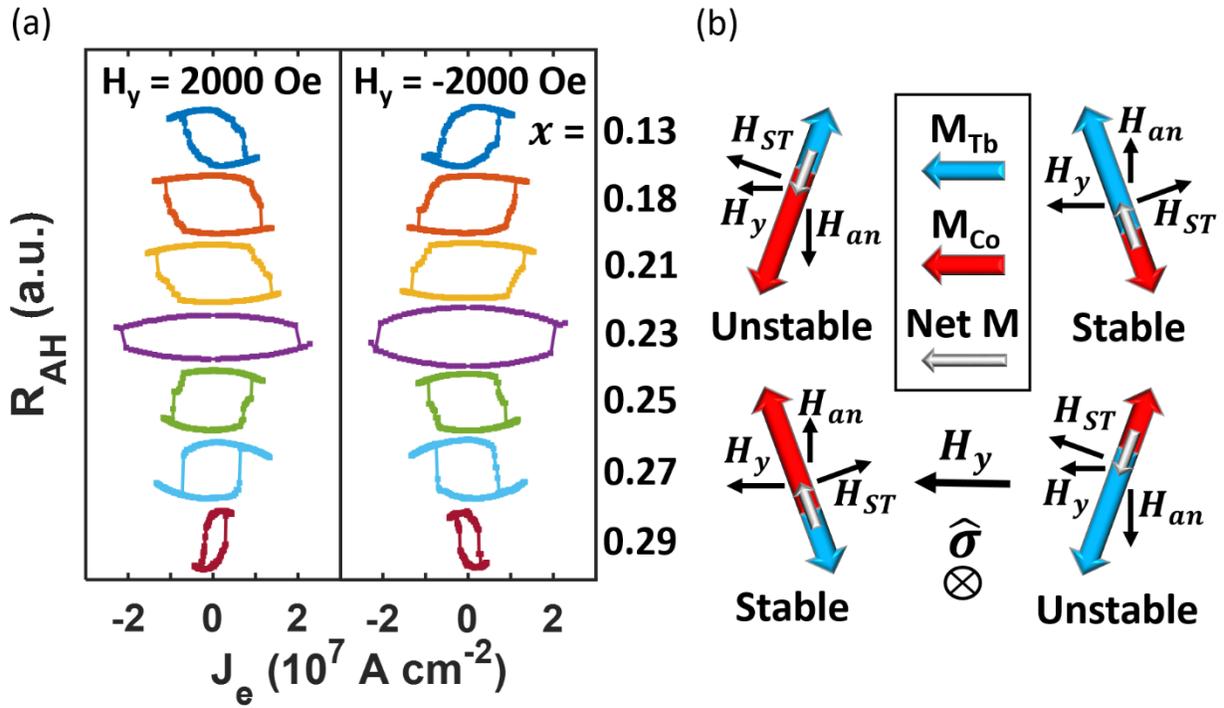

Fig. 2 (a) Current induced SOT switching of $Co_{1-x}Tb_x$ for in-plane fields of $\pm$ 2000 Oe. The current density inside Ta is calculated based on the conductivity of Ta thin films. (b) Schematic of effective fields in a ferrimagnetic system. Fields acting on the moment consist of the in-plane field $H_y$, the anisotropy field $H_{an}$, and the SOT field $H_{ST}$.

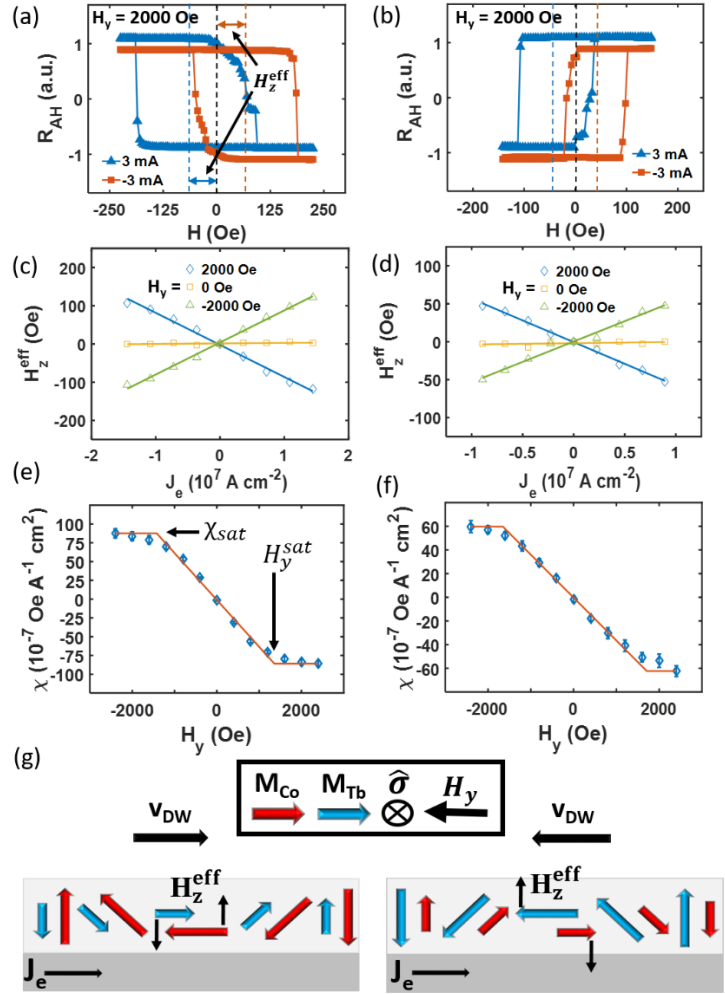

Fig. 3 (a),(c),(e) Measurements on Co-dominant sample $Co_{0.82}Tb_{0.18}$. (b),(d),(f) Measurements on Tb-dominant sample $Co_{0.75}Tb_{0.25}$. (a),(b) $R_{AH}$ vs. applied perpendicular field under a DC current of ±3 mA. (c),(d) SOT effective field as a function of applied current density under in-plane fields of ±2000, and 0 Oe. (e),(f) SOT efficiency vs. $H_y$. Efficiency saturates at the field $H^{sat}_y$. (g) SOT induced DW motion in the ferrimagnetic system for Tb-dominant and Co-dominant films, showing that the effective perpendicular field has the same sign in both cases.

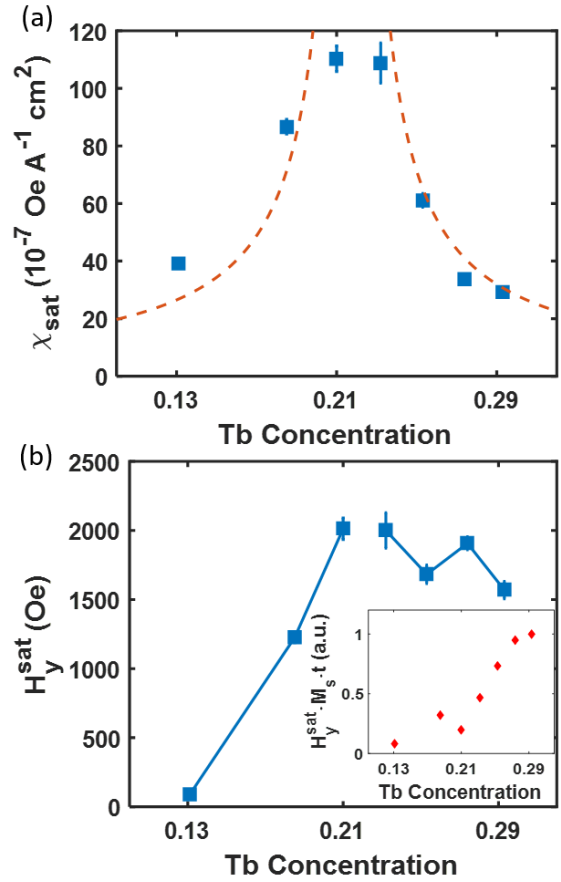

Fig. 4 (a) Saturation efficiency and (b) in-plane saturation field for different $Co_{1-x}Tb_x$ films. Both the saturation efficiency and in-plane saturation field are largest near the magnetic moment compensation point. The dashed line in (a) shows the trend calculated from $\chi_{sat} \sim 1/M_s t$. Inset of (b) plots the product of the in-plane saturation field and magnetic moment $M_s t$, indicating an increase in DMI energy density D with increasing Tb concentration.